\documentclass[12pt, a4paper, final]{iopart}

\newcommand{\SNRT}{\textsc{SNRT}}
\newcommand{\DOS}{\textsc{DOS}}
\newcommand{\MSF}{\textsc{MSF}}
\newcommand{\BS}{\textsc{BS}}
\newcommand{\RT}{\textsc{RT}}
\newcommand{\SNBS}{\textsc{SNBS}}
\newcommand{\GRT}{\textsc{GRT}}

\usepackage{amstext, enumitem, graphicx, iopams}
\expandafter\let\csname equation*\endcsname\relax
\expandafter\let\csname endequation*\endcsname\relax
\usepackage{amsmath}
\usepackage[breaklinks = true, citecolor = blue, colorlinks = true, linkcolor = blue, urlcolor = blue]{hyperref}
\hypersetup{pdfauthor = {Chiranjit Mitra}}

\begin{document}

\title[Recovery time after localized perturbations in complex dynamical networks]{Recovery time after localized perturbations in complex dynamical networks}

\author[cor1]{Chiranjit Mitra$^{1, 2}$, Tim Kittel$^{1, 2}$, Anshul Choudhary$^3$, J\"{u}rgen Kurths$^{1, 2}$, and Reik V. Donner$^1$}
\address{$^1$Research Domain IV - Transdisciplinary Concepts \& Methods, Potsdam Institute for Climate Impact Research, 14473 Potsdam, Germany}
\address{$^2$Department of Physics, Humboldt University of Berlin, 12489 Berlin, Germany}
\address{$^3$Theoretical Physics/Complex Systems, ICBM, Carl-von-Ossietzky University of Oldenburg, 26111 Oldenburg, Germany}
\ead{\mailto{chiranjit.mitra@pik-potsdam.de}}

\begin{abstract}
Maintaining the synchronous motion of dynamical systems interacting on complex networks is often critical to their functionality. However, real-world networked dynamical systems operating synchronously are prone to random perturbations driving the system to arbitrary states within the corresponding basin of attraction, thereby leading to epochs of desynchronized dynamics with a priori unknown durations. Thus, it is highly relevant to have an estimate of the duration of such transient phases before the system returns to synchrony, following a random perturbation to the dynamical state of any particular node of the network. We address this issue here by proposing the framework of \emph{single-node recovery time} (\SNRT{}) which provides an estimate of the relative time scales underlying the transient dynamics of the nodes of a network during its restoration to synchrony. We utilize this in differentiating the particularly \emph{slow} nodes of the network from the relatively \emph{fast} nodes, thus identifying the critical nodes which when perturbed lead to significantly enlarged recovery time of the system before resuming synchronized operation. Further, we reveal explicit relationships between the \SNRT{} values of a network, and its \emph{global relaxation time} when starting all the nodes from random initial conditions. Earlier work on relaxation time generally focused on investigating its dependence on macroscopic topological properties of the respective network. However, we employ the proposed concept for deducing microscopic relationships between topological features of nodes and their respective \SNRT{} values. The framework of \SNRT{} is further extended to a measure of resilience of the different nodes of a networked dynamical system. We demonstrate the potential of \SNRT{} in networks of R\"{o}ssler oscillators on paradigmatic topologies and a model of the power grid of the United Kingdom with second-order Kuramoto-type nodal dynamics illustrating the conceivable practical applicability of the proposed concept.
\end{abstract}

\submitto{\NJP}


\section{\label{sec:Introduction}Introduction}

The abundance of dynamical systems involving large collections of individual entities interacting with each other on complex networks can hardly be further exaggerated~\cite{strogatz2001exploring, albert2002statistical, dorogovtsev2002evolution, newman2003structure, boccaletti2006complex, newman2010networks}. Such networked dynamical systems often exhibit a multitude of stable states, whereby sustained operation of the system in the desired state is of central importance. The desired operational state (\DOS{}) in such systems is commonly associated with the synchronized motion of the dynamical components coupled on their networked architecture~\cite{pikovsky2003synchronization, menck2014dead}. `Permissible' and `impermissible' random perturbations (according to the terminology used by Menck et al.~\cite{menck2013basin}) often disrupt the functionality of coupled dynamical systems operating in the synchronized state, driving them away either to an arbitrary state still inside the basin of attraction of the synchronized state, or to an altogether different dynamical regime. The former situation arising on account of `permissible' perturbations, leads to arbitrary durations of desynchronized dynamics before the system regains synchronous motion. On the other hand, `impermissible' perturbations permanently forbid the return of the system to the synchronized state, unless again affected by an appropriate external perturbation.

The stability of the synchronized state against the aforementioned perturbations is critical in the operation of many real-world networked dynamical systems such as ecosystems, power grids, the human brain, etc.~\cite{mitra2017multiple}. Subsequently, the influence of topological features on network synchronizability and the stability of the synchronized state has been well-investigated~\cite{pikovsky2003synchronization, arenas2008synchronization}. In this context, significant developments constitute the master stability function (\MSF{})~\cite{pecora1998master}, basin stability (\BS{})~\cite{menck2013basin} and its extensions to single-node \BS{}~\cite{menck2014dead}, multiple-node \BS{}~\cite{mitra2017multiple}, and survivability~\cite{hellmann2016survivability}. On the contrary, the issue of recovery time (\RT{}) of complex dynamical networks following a random perturbation, which is a measure of how quickly the network relaxes back to the \DOS{} (e.g., a synchronized state) after being perturbed from the same, has received considerably less attention and is currently under active investigation~\cite{zumdieck2004long, timme2004topological, timme2006speed, qi2008fast, qi2008predicting, son2008relaxation, zillmer2009very, granada2009achieve, grabow2010small, grabow2011speed, wang2016synchronous, kittel2016timing}. However, this is an important problem concerning dynamical robustness of complex networks, i.e., the ability of a network to restore its dynamical activity to the \DOS{} when its dynamical components are subject to random perturbations. For example, the loss of synchrony in engineered systems such as power grids can lead to large-scale power blackouts~\cite{menck2014dead}. In biological systems such as the human brain, it can impede cognitive functions such as information transfer~\cite{fries2005mechanism} and memory~\cite{fell2011role}. Thus, quickly restoring synchrony following desynchronizing perturbations is crucial in such coupled dynamical systems. Consequently, it is highly desirable to have an estimate of the \RT{} of the system to the desired stable regime, following a perturbation to a particular node of the network (otherwise operating in the \DOS{}). This creates the possibility of identifying (and safeguarding) specific nodes which when perturbed lead to particularly large \RT{} of the system. In this regard, we propose here the framework of \emph{single-node recovery time} (\SNRT{}) addressing the aforementioned issue. We reserve a formal definition of \SNRT{} to Section~\ref{sec:SNRT}.

\SNRT{} of a node under investigation relates to the time taken by the system operating in the \DOS{} (e.g., a synchronized state) to return to the same, following a random perturbation to the dynamical state of the respective node. The framework of \SNRT{} provides information on the different relative time scales underlying the transient dynamics of the respective nodes of the network during its restoration to the \DOS{}. This can be utilized in revealing the particularly \emph{slow} nodes of the network in contrast to the relatively \emph{fast} ones, leading to the identification of the vulnerable nodes which when perturbed significantly elevate the \RT{} of the whole system. Further, this can provide an insight into the \emph{global relaxation time} (\GRT{}) of the network to the \DOS{}, when starting all its nodes from arbitrary initial conditions. We provide a formal definition of \GRT{} in Section~\ref{sec:GRT}. The \GRT{} is referred to as the \emph{global synchronization time} when the synchronized state is the \DOS{} of the network.

Previously, the dependence of synchronization time on various macroscopic topological properties of the corresponding networks has been investigated. For example, Grabow et al.~\cite{grabow2010small} have shown that, largely insensitive to the type of oscillators (phase, multi-dimensional, neural), their intrinsic dynamics (periodic, chaotic) and their coupling schemes (phase-difference, diffusive, pulse-like), networks with a fixed average path length consistently synchronize slowest in the small-world regime. This is a rather unexpected phenomenon given that small-world topology has been suggested to facilitate network synchronization at weaker coupling strengths (than for analogous, appropriately normalized globally coupled systems)~\cite{nishikawa2003heterogeneity, watts1998collective, barahona2002synchronization} as well as being more robust to random perturbations~\cite{menck2013basin}. Also, the \MSF{} approach~\cite{pecora1998master} has been extended by Grabow et al.~\cite{grabow2011speed} to provide analytical predictions for the asymptotic synchronization times, which is, however, locally restrictive to small perturbations. Further, the dependence of synchronization time on various macroscopic topological features such as the average path length, global clustering coefficient etc. has been systematically studied. In this context, the framework of \SNRT{} introduced in this work is capable of providing a microscopic view on the response to arbitrary perturbations of individual nodes as well as exploring relationships between various topological features of the nodes and their respective \SNRT{} values.

Finally, we advance on the framework of \SNRT{} for quantifying the resilience of networked dynamical systems~\cite{holling1973resilience}. Resilience of a given dynamical system has been defined in at least two different ways, namely, \emph{engineering resilience} and \emph{ecological resilience}~\cite{holling1996engineering}. Engineering resilience (according to Pimm~\cite{pimm1984complexity}) of a dynamical system characterizes its resistance to disturbance and speed of return to its equilibrium, following a perturbation~\cite{holling1996engineering, gunderson2000ecological, mitra2015integrative}. It implicitly assumes global stability, i.e., the existence of only one equilibrium state, or, if other operating states exist, they should be avoided by applying safeguards~\cite{holling1996engineering, gunderson2000ecological, mitra2015integrative}. On the other hand, ecological resilience~\cite{walker2004resilience} presumes the existence of multiple stable states and the tolerance of the system to disturbances that facilitate transitions among the stable states~\cite{holling1996engineering, gunderson2000ecological, mitra2015integrative}. In this case, resilience of the system is measured by its capacity to remain in the same basin of attraction in the face of random perturbations~\cite{holling1996engineering, gunderson2000ecological, mitra2015integrative}.

Ecological resilience of the multiple stable states of a system relates to the volume and geometry of their respective basins of attraction~\cite{menck2013basin, mitra2015integrative}. In this context, Mitra et al.~\cite{mitra2015integrative} recently reconsidered the concept of ecological resilience and its three crucial aspects of `latitude' ($L$), `resistance' ($R$) and `precariousness' ($Pr$)~\cite{walker2004resilience}. They redefined $L$, $R$ and $Pr$ in a rigorous dynamical systems context and utilized this as a foundation for characterizing multistability and proposing the quantifier of \emph{integral stability}~\cite{mitra2015integrative}. Besides its extension to quantifying multistability, the framework of ecological resilience has generated widespread interest (cf.~\cite{gunderson2012foundations} and references therein). On the other hand, the facet of engineering resilience, perhaps on account of its restrictive scope to globally stable systems has received considerably less attention. However, it is equally crucial to know how long does a system operating in its desired stable state take to retain functionality in the respective dynamical state, following a random perturbation. As mentioned earlier, networked dynamical systems often exhibit multiple stable states, such as the coexistence of synchronized and desynchronized dynamical regimes, which is a notable example of bistable behaviour~\cite{mitra2017multiple}. Thus, we extend here the traditional scope of engineering resilience to quantifying the resilience of the \DOS{} (e.g., the synchronized state) in such multistable coupled dynamical systems. More precisely, we relate the engineering resilience of each node of a networked dynamical system (for the \DOS{}) to the \SNRT{} of the corresponding node such that a node with a lower value of \SNRT{} is considered more resilient and vice versa. Thus, the proposed architecture of engineering resilience complements the existing framework of ecological resilience in characterizing the overall resilience of networked dynamical systems.

This paper is further organized as follows: In Section~\ref{sec:Methods}, we outline the general methodology for calculating \SNRT{} values for a given networked dynamical system. In Section~\ref{sec:Examples}, we illustrate applications of \SNRT{} to networks of R\"{o}ssler oscillators and a model of the power grid of the United Kingdom with second-order Kuramoto-type nodal dynamics. Finally, we present the conclusions of our work in Section~\ref{sec:Conclusions}.


\section{\label{sec:Methods}Methods}


\subsection{\label{sec:Preliminaries}Preliminaries}

Consider a network of $N$ oscillators (nodes) where the intrinsic dynamics of the $i\textsuperscript{th}$ oscillator (represented by the $d$-dimensional state vector $\mathbf{x}_{i}(t) = \left( x_{i}^{1},\, x_{i}^{2},\, \ldots,\, x_{i}^{d} \right)^{\text{T}}$) is described by $\dot{\mathbf{x}}_{i} = \mathbf{F}_{i} \left( \mathbf{x}_{i} \right)$, where $\mathbf{x}_{i} \in \mathbb{R}^{d};\, \mathbf{F}_{i}:\, \mathbb{R}^{d}\, \rightarrow\, \mathbb{R}^{d},\, \mathbf{F}_{i} = \left( F_{i}^{1} \left( \mathbf{x} \right),\, F_{i}^{2} \left( \mathbf{x} \right),\, \ldots,\, F_{i}^{d} \left( \mathbf{x} \right) \right)^{\text{T}};\, i = 1,\, 2,\, \ldots,\, N$. The dynamical equations of the networked system read
\begin{equation} \label{eq:DE_Network}
\dot{\mathbf{x}}_{i} = \mathbf{F}_{i} \left( \mathbf{x}_{i} \right) + \epsilon \sum\limits_{j = 1}^{N} A_{ij} \mathbf{H}_{ij} \left( \mathbf{x}_{i},\, \mathbf{x}_{j} \right),
\end{equation}
where $\epsilon$ is the overall coupling strength, $\mathbf{A}$ is the adjacency matrix which captures the interactions between the nodes such that $A_{ij} \neq 0$ if node $j$ influences node $i$ and $\mathbf{H}_{ij}:\, \mathbb{R}^{d} \times \mathbb{R}^{d}\, \rightarrow\, \mathbb{R}^{d}$ is an arbitrary coupling function from node $j$ to node $i$. For the illustrations in this paper (Sec.~\ref{sec:Examples}), we consider identical nodal dynamics $\left( \mathbf{F}_{i} = \mathbf{F}\, \forall\, i \right)$, symmetric adjacency matrices ($A_{ij} = A_{ji} = 1$ if nodes $i$ and $j$ are connected and $A_{ij} = A_{ji} = 0$ otherwise) and identical coupling functions ($\mathbf{H}_{ij} = \mathbf{H}\, \forall\, i,\, j$).

We assume the desired operational state (\DOS{}) is an attractor of the system that we denote by $\mathcal{A}$ with the corresponding basin of attraction $\mathcal{B} \left( \mathcal{A} \right)$. We usually denote a trajectory on $\mathcal{A}$ by $\tilde{\mathbf{x}}(t)$.


\subsection{\label{sec:RRT}Regularized reaching time}

For a trajectory initiated from $\mathbf{x}(0) = \left( \mathbf{x}_{1}(0),\, \mathbf{x}_{2}(0),\, \ldots,\, \mathbf{x}_{N}(0) \right)^{\text{T}}$ ($\in \mathcal{B} \left( \mathcal{A} \right)$), the attractor is usually reached asymptotically. This implies that the associated reaching time is not finite, thus posing a problem in its measurement. A way to address this problem is regularization of the time variable~\cite{kittel2016timing}. We now discuss the framework of \emph{regularized reaching time} proposed by Kittel et al.~\cite{kittel2016timing} and then resort to the same in dealing with the above issue.

The distance of a state at time $t$ on a trajectory initiated from $\mathbf{x}(0)$, to the desired attractor is given by,
\begin{equation*}
d \left( \mathbf{x} \left( t,\, \mathbf{x}(0) \right), \mathcal{A} \right) = \text{inf}\, \{ \| \mathbf{x} \left( t,\, \mathbf{x}(0) \right) - \mathbf{x}^{\prime} \|,\, \forall\, \mathbf{x}^{\prime} \in \mathcal{A} \},
\end{equation*}
where $\mathbf{x} \left( t,\, \mathbf{x}(0) \right)$ represents the state of the system after a time $t$ has elapsed. The last-entry time for the corresponding trajectory to enter a $\delta$-neighbourhood around the desired attractor $\left( \mathcal{A} \right)$ is given by
\begin{equation*}
t_{L} \left( \mathbf{x}(0),\, \delta \right) = \text{inf}\, \{ T: d \left( \mathbf{x} \left( t,\, \mathbf{x}(0) \right), \mathcal{A} \right) < \delta,\, \forall\, t \geq\, T \},
\end{equation*}
where $\delta \rightarrow 0$ leads to the aforementioned divergence.

Kittel et al.~\cite{kittel2016timing} argued that even though the actual reaching times diverge for the respective trajectories, their differences actually converge. Subsequently, they proposed the \emph{regularized reaching time} $T_{RR} \left( \mathbf{x}(0) \right)$ for any trajectory (starting from $\mathbf{x}(0)$) as the difference between the last-entry times along the respective trajectory and a reference trajectory (starting from $\mathbf{x}_{ref}$), for a given $\delta > 0$. This can be interpreted as the additional time the trajectory starting from $\mathbf{x}(0)$ needs to arrive in the vicinity of the desired attractor, after the reference trajectory starting from $\mathbf{x}_{ref}$ has reached it. Thus,
\begin{equation} \label{eq:T_RR}
T_{RR} \left( \mathbf{x}(0) \right) = \lim\limits_{\delta \rightarrow 0} \left( t_{L} \left( \mathbf{x}(0),\, \delta \right) - t_{L} \left( \mathbf{x}_{ref},\, \delta \right) \right).
\end{equation}
A positive or negative value of $T_{RR} \left( \mathbf{x}(0) \right)$ indicates that the considered trajectory arrives by this value \emph{later} or \emph{earlier} than the reference trajectory, respectively. This allows the distinction between \emph{slower} and \emph{faster} trajectories of the system during their return to the desired attractor (cf. Kittel et al.~\cite{kittel2016timing} for further details on $T_{RR}$).


\subsection{\label{sec:SNRT}Single-node recovery time (SNRT)}

In the following, we outline the general methodology for calculating \SNRT{} values for all nodes of any networked dynamical system. We assume that the networked dynamical system of Eq.~(\ref{eq:DE_Network}) is in its \DOS{} $\mathbf{\tilde{x}}(t)$. Now, consider a `permissible' random perturbation $\Delta \mathbf{x}_{i}$ to the dynamical state of the $i\textsuperscript{th}$ oscillator of the network. The system (otherwise functioning in its \DOS{}) is pushed to a perturbed state $\mathbf{x}_{\Delta i} = \left( \mathbf{\tilde{x}}_{1},\, \mathbf{\tilde{x}}_{2},\, \ldots,\, \mathbf{\tilde{x}}_{i} + \Delta \mathbf{x}_{i},\, \ldots,\, \mathbf{\tilde{x}}_{N} \right)^{\text{T}}$. The perturbed state (on account of the perturbation being `permissible') remains in the basin of attraction $\mathcal{B} \left( \mathcal{A} \right)$ of the \DOS{} because we chose $ \mathbf{x}_{\Delta i}$ to be permissible, thus ensuring the system's return to the same. We then define the \SNRT{} of the $i\textsuperscript{th}$ oscillator as,
\begin{equation} \label{eq:SNRT}
\langle T_{R}^{1} \left( i \right) \rangle = \frac{\displaystyle \int\limits_{P_i(\mathcal{B} \left( \mathcal{A} \right))} \rho_{i} \left( \mathbf{x}_{\Delta i} \right)\, T_{RR} \left( \mathbf{x}_{\Delta i} \right)\, d \Delta \mathbf{x}_{i}}{\displaystyle \int\limits_{P_i(\mathcal{B} \left( \mathcal{A} \right))} \rho_{i} \left( \mathbf{x}_{\Delta i} \right)\, d \Delta \mathbf{x}_{i}},
\end{equation}
where $P_i$ is the projector into the subspace of the $i^\text{th}$ oscillator, i.e., \ $P_i(\mathbf{x}) = \mathbf{x}_i$. $\rho_{i} \left( \mathbf{x}_{\Delta i} \right)$ is the density of `permissible' perturbed states in state space that the $i\textsuperscript{th}$ oscillator may be pushed to even via large perturbations with $\displaystyle \int \rho_{i} \left( \mathbf{x}_{\Delta i} \right)\, d \Delta \mathbf{x}_{i} = 1$, where this integral is performed over the subspace of the $i\textsuperscript{th}$ oscillator. The integrals in the numerator and denominator of Eq.~(\ref{eq:SNRT}) are performed over the basin of attraction of the \DOS{} $\left( \text{i.e.,}\ \mathbf{x}_{\Delta i} \in \mathcal{B} \left( \mathcal{A} \right) \right)$. Thus, the \SNRT{} of the $i\textsuperscript{th}$ oscillator $\langle T_{R}^{1} \left( i \right) \rangle$ corresponds to the mean regularized reaching time of the system to the \DOS{}, after a random `permissible' perturbation hits the respective oscillator.

Equation~\eqref{eq:T_RR} demands the choice of a reference initial condition $\mathbf{x}_{ref} \in \mathcal{B} \left( \mathcal{A} \right)$ that needs to be kept fixed for all single-node perturbations to allow comparability between \SNRT{} values of the different nodes of the network. However, different choices of $\mathbf{x}_{ref}$ (as long as we do not choose it on $\mathcal{A}$) simply lead to a shift of all $\langle T_{R}^{1} \left( i \right) \rangle$ values by a constant only~\cite{kittel2016timing}. Although not posing a serious problem, this methodology of choosing $\mathbf{x}_{ref}$ leaves an element of arbitrariness. As we seek to utilize the $\langle T_{R}^{1} \rangle$ values in estimating the duration by which a particular node of the network returns \emph{faster} or \emph{slower} than another, this naturally leads to the condition demanding the lowest $\langle T_{R}^{1} \left( i \right) \rangle$ value to be $0$,
\begin{equation} \label{eq:x_ref}
\min_{i} \left( \langle T_{R}^{1} \left( i \right) \rangle \right) = \langle T_{R}^{1} \rangle_{min} = 0.
\end{equation}
Using this equation, we can fix $\mathbf{x}_{ref}$ implicitly instead of explicitly specifying it. We denote the node (or one representative if there might be more) with $\langle T_{R}^{1} \left( i \right) \rangle = 0$ by $i_{ref}$. The resulting values of $\langle T_{R}^{1} \rangle$ now represent differences in time by which nodes of the network return \emph{faster} or \emph{slower} than the reference node $i_{ref}$. As opposed to arbitrarily choosing $\mathbf{x}_{ref}$, thereby resulting in negative $T_{RR}$ values (which is counter-intuitive when measuring time), the above choice of $\mathbf{x}_{ref}$ ensures non-negativity of $\langle T_{R}^{1} \left( i \right) \rangle$ values, besides eliminating the arbitrariness associated with the choice of $\mathbf{x}_{ref}$. Further details on the choice of the reference trajectory are provided in~\ref{sec:Appendix_x_ref}.

We now present an algorithm for estimating the \SNRT{} of the $i\textsuperscript{th}$ oscillator/node of a network (modelled using Eq.~(\ref{eq:DE_Network})):
\begin{enumerate}[label=(\roman*)]
\item Identify the \DOS{} of the network. This state often corresponds to the synchronized dynamics of the oscillators coupled on the network.
\item When the attractor corresponding to the \DOS{} $\mathcal{A}$ is not a fixed point, choose $P\ \left( > 1 \right)$ different points on the attractor. Otherwise, choose $P = 1$.
\item For a particular value of $p\, \left( p = 1,\, 2,\, \ldots, \, P \right)$, perturb the $i\textsuperscript{th}$ oscillator by drawing $I_{C}$ randomly distributed $\left(\text{according to}\ \rho_{i} \left( \mathbf{x}_{\Delta i} \right) \right)$ initial conditions $\mathbf{x}_{\Delta i}^{p} \left( j = 1,\, 2,\, \ldots, \, I_{C} \right)$ from inside the basin of attraction of the \DOS{}, while each time initiating the system from the \DOS{} corresponding to the $p\textsuperscript{th}$ point on $\mathcal{A}$. For the results described in this paper, we assume a uniform distribution of $\rho_{i} \left( \mathbf{x}_{\Delta i} \right)$.
\item For a fixed value of $\delta > 0$, calculate the last-entry time $t_{L} \left( \mathbf{x}_{\Delta i}^{p} \left( j \right),\, \delta \right)$ of the system for the $j\textsuperscript{th}$ initial condition.
\item Estimate the \SNRT{} of the $i\textsuperscript{th}$ oscillator $\left( T_{R}^{1} \left( i,\, p \right) \right)$ for the $p\textsuperscript{th}$ point on the attractor as,
\begin{equation}
\hat{T}_{R}^{1} \left( i,\, p \right) = \frac{\sum\limits_{j = 1}^{I_{C}} t_{L} \left( \mathbf{x}_{\Delta i}^{p} \left( j \right),\, \delta \right)}{I_{C}},
\end{equation}
and then average over $p$ to obtain,
\begin{equation}
\langle T_{R}^{1} \left( i \right) \rangle = \frac{1}{P} \sum\limits_{p = 1}^{P} \hat{T}_{R}^{1} \left( i,\, p \right).
\end{equation}
\item Finally, we identify the node $i_{ref}$ with the minimum $\langle T_{R}^{1} \rangle$ value (as computed above for all nodes) $\langle T_{R}^{1} \rangle_{min}$ and subtract this value from the $\langle T_{R}^{1} \left( i \right) \rangle$ of the $i\textsuperscript{th}$ oscillator computed above, thus yielding the \SNRT{} value of the respective oscillator.
\end{enumerate}

As mentioned earlier, this concept of \SNRT{} can be utilized in identifying the \emph{slow} and \emph{fast} nodes/sub-components of networked dynamical systems. Also, the proposed machinery can be used in revealing systematic relationships between \SNRT{} values of different nodes and their respective topological features. Further, it can be extended to a measure of (engineering) resilience of the different nodes of a networked dynamical system (see Sec.~\ref{sec:ER}) and thereby utilized in identifying the particularly vulnerable nodes of the network as well as the more resilient ones. Subsequently, this framework of \SNRT{} can be potentially relevant in selecting specific nodes to be safeguarded from external perturbations. We now define the \emph{global relaxation time} of a network, which relates to the overall time scale of the dynamics of a network during its relaxation to the \DOS{}.


\subsection{\label{sec:GRT}Global relaxation time (GRT)}

Starting all nodes of a networked dynamical system from random initial conditions inside the basin of attraction of the desired attractor involves a transient time before the system reaches the associated attractor. We label the duration of this transient regime as the \emph{relaxation time} of the system for the respective initial state. We estimate the \emph{global relaxation time} $\langle T_{R} \rangle$ of a network as follows:
\begin{enumerate}[label=(\roman*)]
\item Draw $I_{C}$ random initial conditions from inside the basin of attraction of the \DOS{}. The $j\textsuperscript{th}$ initial condition can be written as $\mathbf{x} \left( j \right) = \left( \mathbf{x}^{1},\, \mathbf{x}^{2},\, \ldots,\, \mathbf{x}^{N} \right)^{\text{T}}$ where $j = 1,\, 2,\, \ldots, \, I_{C}$. Note, that the value of $I_{C}$ chosen for computing the \GRT{} can be different from the one chosen for calculating \SNRT{} above (Sec.~\ref{sec:SNRT}).
\item For the $j\textsuperscript{th}$ initial condition, calculate the last-entry time $t_{L} \left( \mathbf{x} \left( j \right),\, \delta \right)$ of the system with the same value of $\delta$ as chosen for computing \SNRT{} (Sec.~\ref{sec:SNRT}).
\item Calculate the \GRT{} of the network as,
\begin{equation}
\langle T_{R} \rangle = \frac{1}{I_{C}} \sum\limits_{j = 1}^{I_{C}} t_{L} \left( \mathbf{x} \left( j \right),\, \delta \right).
\end{equation}
\item Finally, subtract the value of $\langle T_{R}^{1} \rangle_{min}$ (obtained in Sec.~\ref{sec:SNRT}) from the $\langle T_{R} \rangle$ computed above in obtaining the \GRT{} of the network.
\end{enumerate}
When the \DOS{} of the network is a synchronized state, its \GRT{} is referred to as the \emph{global synchronization time} of the system.

The \GRT{} of a network is useful for quantifying the expected transient time to reach the \DOS{}, when starting the system from a random initial condition. In Section~\ref{sec:Examples}, we will illustrate the relationship between \SNRT{} values and the \GRT{} of a network for different systems.

In order to avoid terminological confusion, we explicitly distinguish between the usage of \emph{recovery}, \emph{reaching} and \emph{relaxation} time. We use the term \emph{recovery} with reference to the time taken by the system to recover from a perturbation and resume operation in the \DOS{}. On the other hand, when initiating all the nodes of the system from arbitrary conditions, the term \emph{relaxation} is used with reference to the time before the system relaxes to the DOS{}. It is the difference between the relaxation times of a trajectory starting from a particular initial condition and that of a reference trajectory, which is termed as the regularized \emph{reaching} time for the respective initial condition.


\subsection{\label{sec:SNBS}Single-node basin stability (SNBS)}

The \BS{} of a particular attractor relates the volume of its basin of attraction to the likelihood of returning to the same attractor in the face of random perturbations~\cite{menck2014dead, menck2013basin, mitra2017multiple}. More precisely, the \BS{} of a particular attractor is defined as the fraction of the volume of the state space belonging to the basin of attraction of the respective attractor~\cite{menck2014dead, menck2013basin, mitra2017multiple}. In practice, \BS{} of any particular attractor is estimated using a numerical Monte-Carlo procedure by drawing random initial states from a chosen subset of the entire state space, simulating the associated trajectories, and calculating the fraction of trajectories that approach the respective attractor~\cite{menck2014dead, menck2013basin, mitra2017multiple}. As mentioned earlier, the ecological resilience of a stable state is (among other properties) determined by the size and shape of its basin of attraction, and is therefore closely related to its \BS{}.

\BS{} has been further extended to the framework of single-node \BS{} (\SNBS{})~\cite{menck2014dead, mitra2017multiple}. \SNBS{} $\langle S_{B}^{1} \rangle$ of a node under investigation corresponds to the probability of the network (operating in the \DOS{}) to return to the \DOS{}, after that particular node has been hit by a non-infinitesimal perturbation~\cite{menck2014dead, mitra2017multiple}. We refer to Mitra et al.~\cite{mitra2017multiple} for the general methodology used throughout this paper for estimating \SNBS{} values for any networked dynamical system.


\subsection{\label{sec:ER}Engineering resilience}

\SNBS{} is a measure related to the ecological resilience of a node subjected to a random perturbation (when the entire network was functioning in the \DOS{} prior to the disturbance). The time elapsed before the network returns to its \DOS{}, following a `permissible' random perturbation to a particular node determines the engineering resilience of the respective node. We recommend incorporating the engineering resilience of a node (besides its ecological resilience as characterized by its \SNBS{} value) quantified as being inversely related to its \SNRT{} value, in measuring the overall resilience of the respective node. For example, it may be possible that two nodes of a networked dynamical system have very similar values of \SNBS{}. However, the \SNRT{} values of the respective nodes may differ significantly (as we shall illustrate using examples in Sec.~\ref{sec:Examples}). In such a situation, the new framework of \SNRT{} should complement that of \SNBS{} in appropriately assessing the resilience of the respective nodes of a network.


\section{\label{sec:Examples}Examples}

We shall now illustrate applications of \SNRT{} to various networked dynamical systems. Here, we specifically apply the framework to networks of oscillators with continuous time dynamics (Eq.~(\ref{eq:DE_Network})) exhibiting bistability on account of coexisting synchronized and desynchronized regimes, where the former is considered as the \DOS{} of the system. However, the framework is generally applicable to (continuous or discrete time) networked dynamical systems with multiple coexisting states as well.


\subsection{\label{sec:DSFN_RO}Deterministic scale-free network of R\"{o}ssler oscillators}

\begin{figure*} [h]
\begin{center}
\includegraphics[height=8.0cm, width=18.0cm]{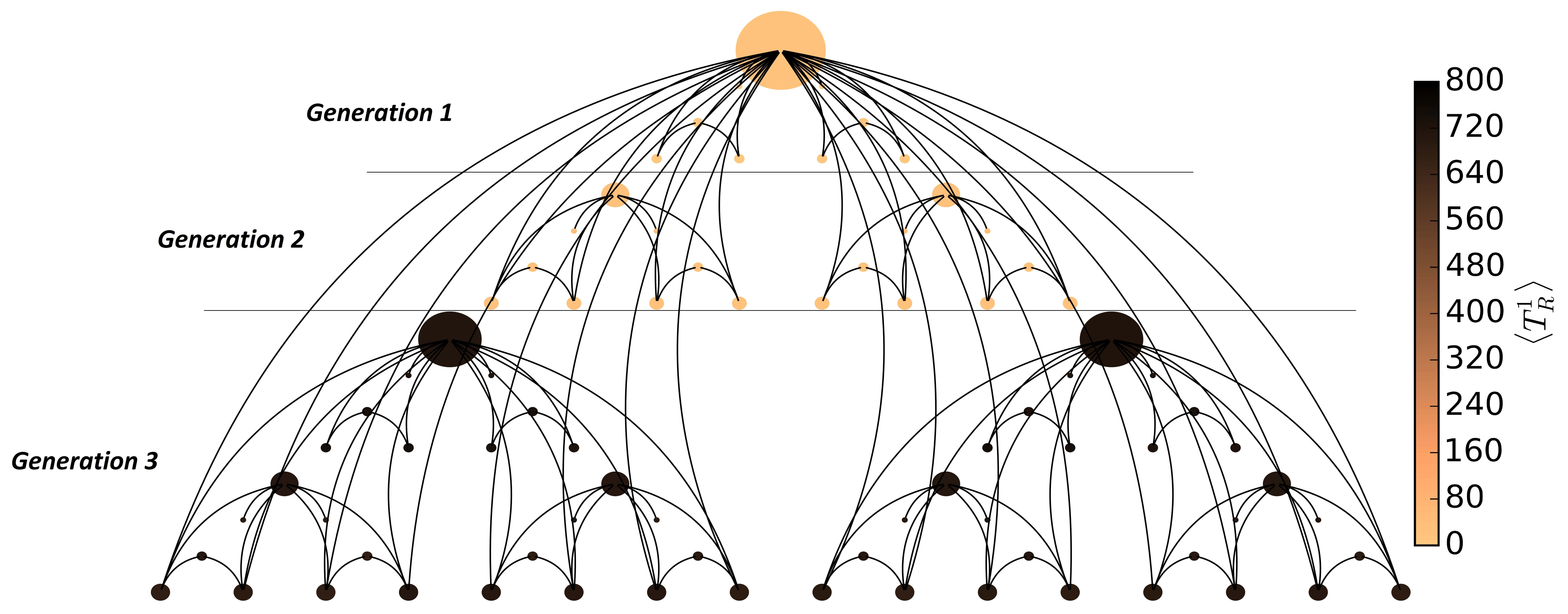}
\caption{\label{fig:Figure_1}(Color online) Network topology of the undirected deterministic scale-free network of $N = 81$ identical R\"{o}ssler oscillators. The size of each node is proportional to its degree and the color indicates the $\langle T_{R}^{1} \rangle$ value of the respective node.}
\end{center}
\end{figure*}

We first consider a network of $N$ identical R\"{o}ssler oscillators~\cite{rossler1976equation}, with diffusive coupling in the $y$-variable between two coupled nodes such that the full dynamical equations of node $i$ (in analogy with Eq.~(\ref{eq:DE_Network})) read
\begin{equation} \label{eq:DSFN_RO_DE}
\begin{split}
\dot{x}_{i}^{1} & = - x_{i}^{2} - x_{i}^{3},\\
\dot{x}_{i}^{2} & = x_{i}^{2} + a x_{i}^{2} + \epsilon \sum\limits_{j = 1}^{N} A_{ij} \left( x_{j}^{2} - x_{i}^{2} \right),\\
\dot{x}_{i}^{3} & = b + x_{i}^{3} \left( x_{i}^{1} - c \right).
\end{split}
\end{equation}
We use the parameter values of $a = b = 0.2$ and $c = 7.0$ for which the intrinsic dynamics of each uncoupled R\"{o}ssler oscillator is chaotic.

As a specific network topology, we use an undirected deterministic scale-free network proposed by Barab\'{a}si, Ravasz and Vicsek~\cite{barabasi2001deterministic}. For the simulations carried out in this section, we generate a deterministic scale-free network developed over 3 generations and hence, comprising $N = 81$ nodes (Fig.~\ref{fig:Figure_1}).

\begin{figure*}
\begin{center}
\includegraphics[height=8.0cm, width=18.0cm]{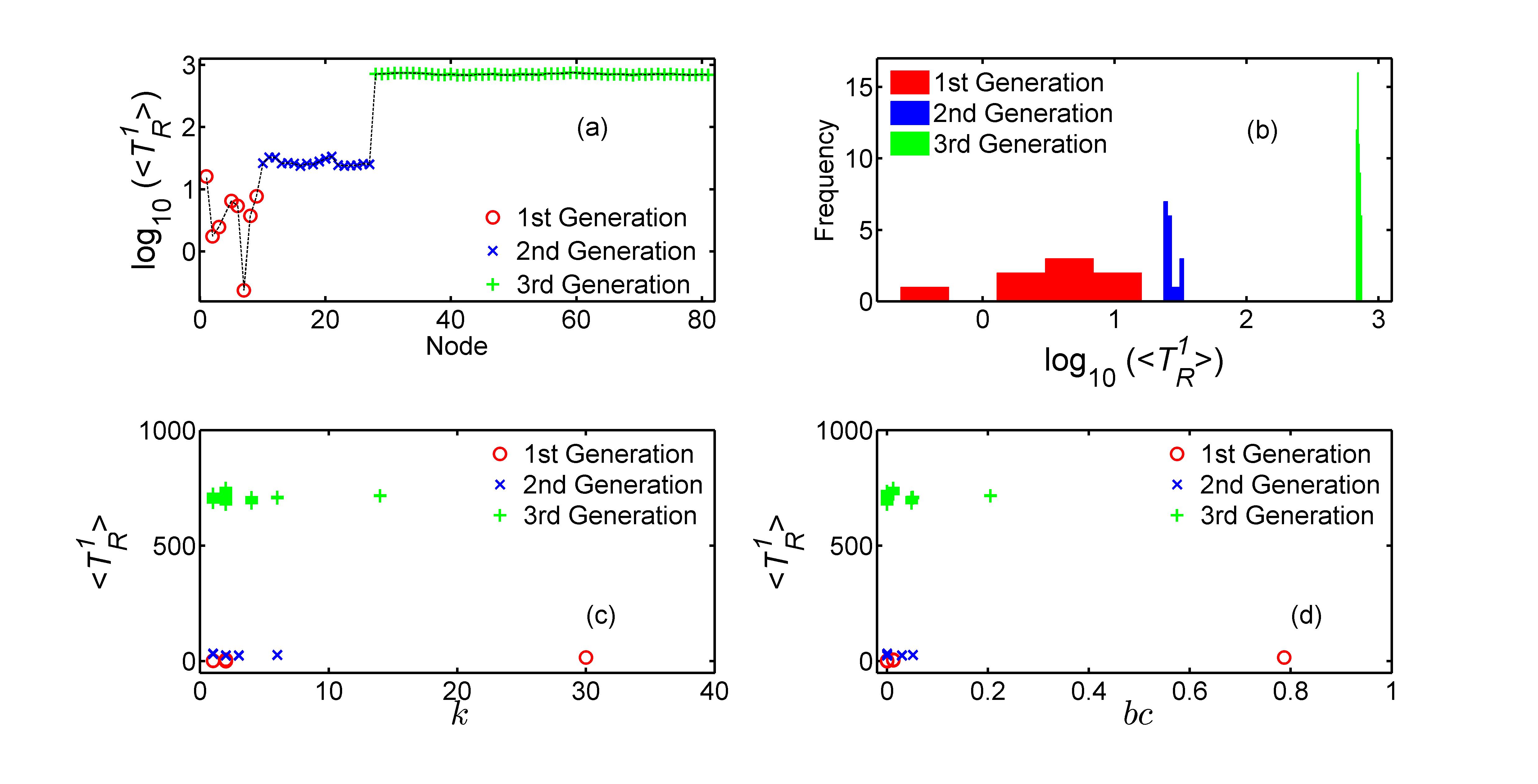}
\caption{\label{fig:Figure_2}(Color online) (a) \SNRT{} $\langle T_{R}^{1} \rangle$ (on $\log_{10}$ scale) of the nodes of the 3 generations of the undirected deterministic scale-free network of $N$ identical R\"{o}ssler oscillators (Eq.~(\ref{eq:DSFN_RO_DE})). The first 9 nodes comprise the $1\textsuperscript{st}$ generation, the next 18 nodes the $2\textsuperscript{nd}$ generation and the final 54 nodes the $3\textsuperscript{rd}$ generation. Node $4$ having the minimum \SNRT{} value $\langle T_{R}^{1} \left( 4 \right) \rangle = 0$ of the network (implying divergence of $\log_{10} \left( \langle T_{R}^{1} (4) \rangle \right)$) has not been shown in the plot. (b) Histogram of $\log_{10} \left( \langle T_{R}^{1} \rangle \right)$ of the nodes. (c, d) Relationship of $\langle T_{R}^{1} \rangle$ with (c) degree ($k$) and (d) betweenness centrality ($bc$) of the nodes.}
\end{center}
\end{figure*}

We consider the \DOS{} of the network as the completely synchronized state, which corresponds to all oscillators following the same trajectory. Further, we choose $\epsilon = 0.8$ from the stability interval (of the completely synchronized state) predicted by the \MSF{}~\cite{pecora1998master} and set $\delta = 10^{-4}$~\cite{note_choice_of_delta} for estimating the \SNRT{} $\left( \langle T_{R}^{1} \rangle \right)$ values, using the procedure described in Section~\ref{sec:SNRT}.

We calculate and present the individual $\langle T_{R}^{1} \rangle$ (on $\log_{10}$ scale) values of the nodes in Fig.~\ref{fig:Figure_2}(a). Interestingly, the 3 generations of nodes split into three classes in terms of their $\langle T_{R}^{1} \rangle$ values such that the lower the generation in the hierarchy, the higher is the \SNRT{} of the individual nodes comprising it (as evident from the histogram in Fig.~\ref{fig:Figure_2}(b)). We next compare these findings with two key topological features of DSF network. The connectivity of a node $i$ (for $i = 1,\, 2,\, \ldots,\, 81$) is described by its degree $k_{i} = \sum\limits_{j} A_{ij}$ (where $\mathbf{A}$ is again the adjacency matrix of the respective network~\cite{newman2010networks}). The betweenness centrality $bc_{i}$ of a node $i$ is related to the fraction of shortest paths between all pairs of nodes that pass through node $i$~\cite{newman2010networks}. For an $N$-node network, the betweenness centrality of each node may further be normalized by dividing by the number of node pairs excluding $N$ $\left( \text{i.e.},\ {N \choose 2} \right)$, obtaining a value between 0 and 1. Thus, $bc_{i} = \frac{2}{N \left( N - 1 \right)} \sum\limits_{j \neq k \neq i} \frac{\sigma_{j, k}^{i}}{\sigma_{j, k}}$ where $\sigma_{j, k}$ is the total number of shortest paths from node $j$ to node $k$ and $\sigma_{j, k}^{i}$ is the number of such shortest paths which pass through node $i$~\cite{newman2010networks}. Figure~\ref{fig:Figure_2}(c, d) shows the relationship of the $\log_{10} \left( \langle T_{R}^{1} \rangle \right)$ values with the topological features of degree $k$ and betweenness centrality $bc$ of the nodes, respectively. The $\langle T_{R}^{1} \rangle$ values do not exhibit any marked relationship with these two characteristics. This is further illustrated by the correlation coefficient of -0.040 (-0.085) between $\langle T_{R}^{1} \rangle$ and $k\, \left( bc \right)$. We summarize our results in Fig.~\ref{fig:Figure_1}, which displays the network topology where the size of each node is proportional to the degree and the color corresponds to the $\langle T_{R}^{1} \rangle$ value of the respective node.

The nodes in the $3\textsuperscript{rd}$ generation of the deterministic scale-free network comprise its \emph{slow} nodes. It is expected that the overall time scale of synchronization of a network should be governed by the node with the highest \SNRT{}, i.e., the `slowest' node of the system. The `slowest' node of the deterministic scale-free network has $\langle T_{R}^{1} \rangle \approx 749.8$. We also computed the \GRT{} $\langle T_{R} \rangle$ of the deterministic scale-free network using the methodology described in Section~\ref{sec:GRT}. We find $\langle T_{R} \rangle \approx 750.04$ being very close to the maximum $\langle T_{R}^{1} \rangle$ value of the network. Thus, we conclude that the `slowest' nodes of the deterministic scale-free network indeed govern its overall time scale of synchronization. However, this result cannot be generalized to any arbitrary topology as we will demonstrate in the following.


\subsection{\label{sec:RSFN_RO}Random scale-free networks of R\"{o}ssler oscillators}

\begin{figure*}
\begin{center}
\includegraphics[height=8.0cm, width=18.0cm]{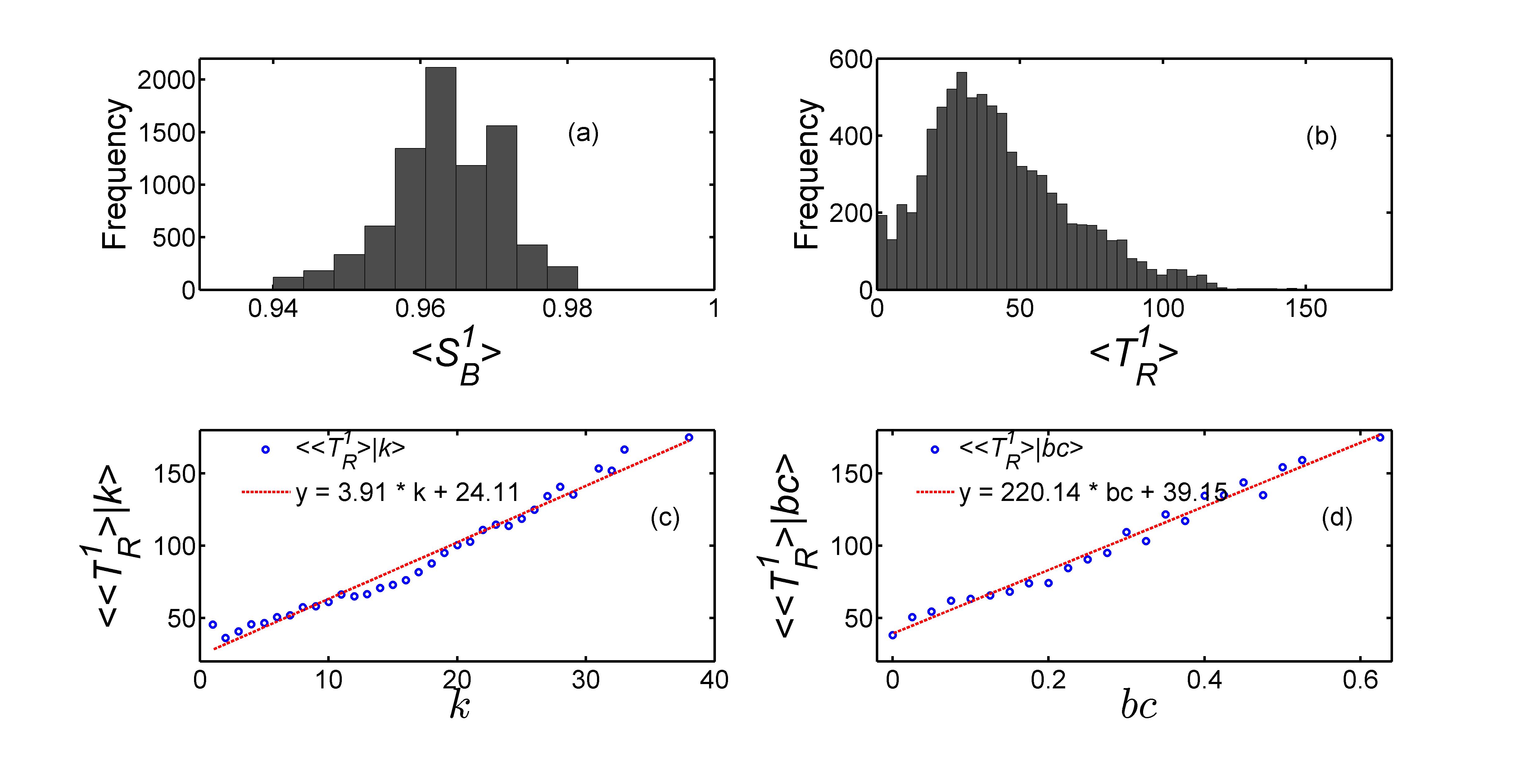}
\caption{\label{fig:Figure_3}(Color online) (a) Histogram of \SNBS{} $\langle S_{B}^{1} \rangle$ of all nodes of the considered ensemble of random scale-free networks. (b) Same for the $\langle T_{R}^{1} \rangle$ values. (c, d) Conditional means (blue circles) of $\langle T_{R}^{1} \rangle$ with respect to (c) degree $\left( \langle \langle T_{R}^{1} \rangle \mid k \rangle \right)$ and (d) betweenness centrality $\left( \langle \langle T_{R}^{1} \rangle \mid bc \rangle \right)$ of the nodes. The red lines indicate linear fits to the conditional means.}
\end{center}
\end{figure*}

\begin{figure*}
\begin{center}
\includegraphics[height=8.0cm, width=18.0cm]{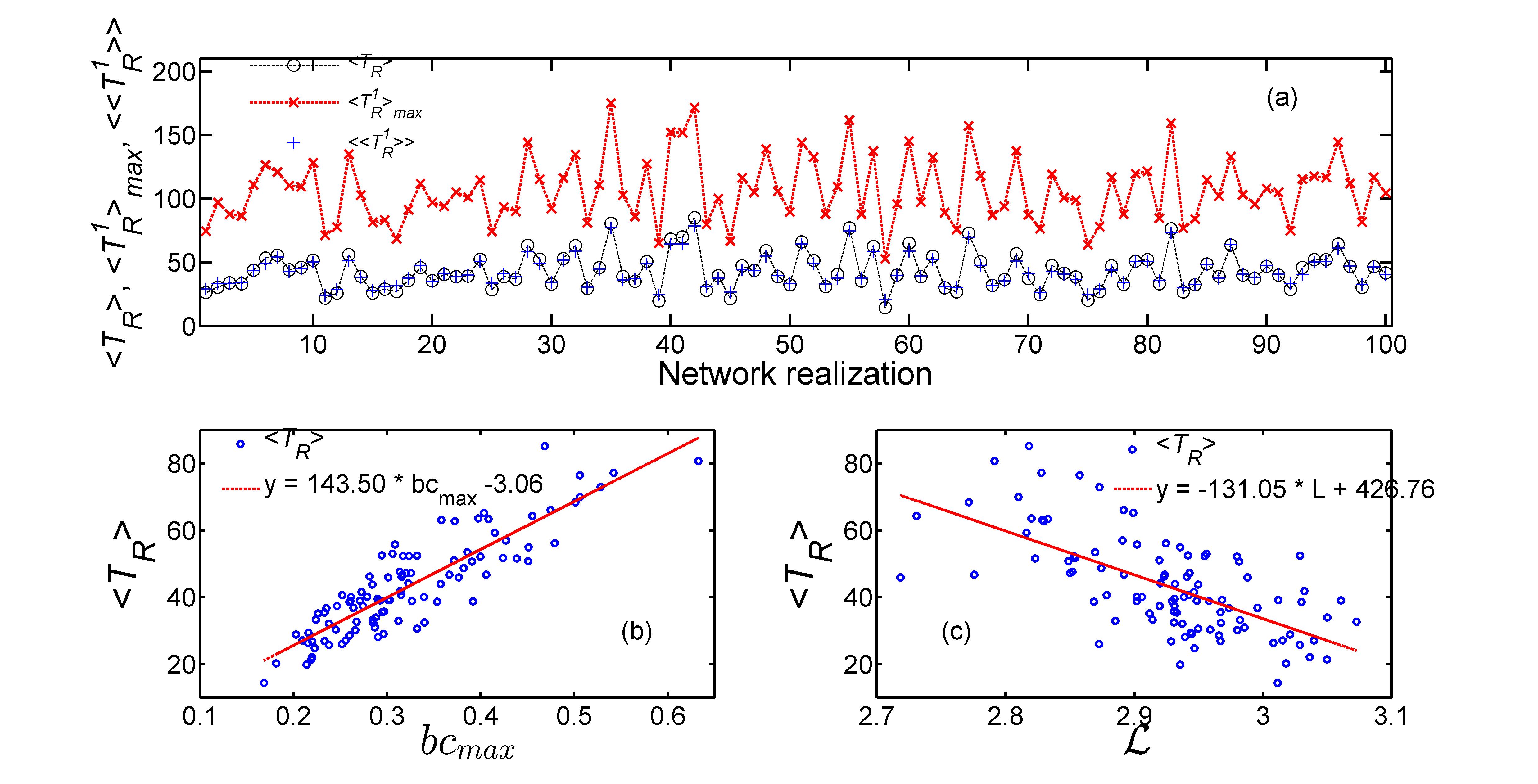}
\caption{\label{fig:Figure_4}(Color online) (a) Global \RT{} $\langle T_{R} \left( \tilde{x} \right) \rangle$ (black circles), maximum \SNRT{} $\langle T_{R}^{1} \left( \tilde{x} \right) \rangle_{max}$ (red crosses) and average \SNRT{} $\langle \langle T_{R}^{1} \left( \tilde{x} \right) \rangle \rangle$ (blue crosses) of all network realizations from the considered ensemble of random scale-free networks. (b) Relationship between $\langle T_{R} \left( \tilde{x} \right) \rangle$ (blue circles) and the maximum betweenness centrality $\left( bc_{max} \right)$ of all nodes of the respective network realization. (c) As in (b) for $\langle T_{R} \left( \tilde{x} \right) \rangle$ and average path length $\left( \mathcal{L} \right)$ of the respective network realization. The red lines in (b, c) indicate linear fits.}
\end{center}
\end{figure*}

Next, we consider an ensemble of 100 random scale-free networks (generated using the classical Barab\'{a}si-Albert (BA) model of growth and preferential attachment~\cite{barabasi1999emergence}) of $N = 81$ R\"{o}ssler oscillators each, with the same parameter values as for the deterministic scale-free network. While generating the random scale-free networks, we explicitly model the growing character of the network by starting with a small number of vertices and at every time step introducing a new vertex and linking it to 2 vertices already present in the system (until the network comprises 81 nodes). Preferential attachment is incorporated by assuming that the probability $\Pi_{i}$ that a new node will be connected to node $i$ depends on the degree $k_{i}$ of node $i$, such that $\Pi_{i} = \frac{k_{i}}{\sum\limits_{j} k_{j}}$. The deterministic scale-free network of $N = 81$ R\"{o}ssler oscillators studied in Section~\ref{sec:DSFN_RO} has 130 edges, equivalently, an edge density of $\frac{130}{{81 \choose 2}} \approx 0.04$. The random scale-free networks generated using the classical BA model have edge densities (similar to that of the deterministic scale-free network) of 0.049, i.e., 158 edges in each realization. Therefore, the results obtained for both topologies are not directly comparable quantitatively.

The distribution of \SNBS{} $\langle S_{B}^{1} \rangle$ values of the $N = 81$ nodes of the considered ensemble is presented in Fig.~\ref{fig:Figure_3}(a). Surprisingly, all nodes have similar and very high $\langle S_{B}^{1} \rangle$ values. Similar results have been observed in a recent study on \SNBS{} values in the deterministic scale-free network of R\"{o}ssler oscillators~\cite{mitra2017multiple}. These observations lead to two important conclusions. Firstly, the similar and rather high $\langle S_{B}^{1} \rangle$ values indicate that the synchronized state in scale-free networks is generally very robust to perturbations affecting a single node of the system. Secondly, we observe that the presence or lack of such a specific macroscopic (hierarchical) structure in the respective scale-free network does not affect the distribution of its $\langle S_{B}^{1} \rangle$ values markedly. In contrast to the latter finding, we have already observed an influence of the hierarchical structure on $\langle T_{R}^{1} \rangle$ for the deterministic scale-free network (Fig.~\ref{fig:Figure_2}(a)). On this note, we shall further unfold dependences of $\langle T_{R}^{1} \rangle$ values on different topological features of random scale-free networks.

The corresponding distribution of $\langle T_{R}^{1} \rangle$ (for $\delta = 10^{-6}$~\cite{note_choice_of_delta}) of all nodes of the considered ensemble of random scale-free networks is shown in Fig.~\ref{fig:Figure_3}(b). As in the case of the deterministic scale-free network, we next consider the mutual dependence between \SNRT{} and the local topological characteristics of the network. For this purpose, we study the distribution of $\langle T_{R}^{1} \rangle$ values of all nodes of the ensemble with respect to their degree and betweenness centrality. We collect all nodes of the ensemble having a particular degree $k$ and calculate the mean over the $\langle T_{R}^{1} \rangle$ values of all these nodes which corresponds to the conditional mean $\langle \langle T_{R}^{1} \rangle \mid k \rangle$. Similarly, we bin the $bc$ values of all nodes of the ensemble and calculate the conditional mean $\langle \langle T_{R}^{1} \rangle \mid bc \rangle$ over the $\langle T_{R}^{1} \rangle$ values of all nodes belonging to the respective bin. Interestingly, the conditional mean values exhibit a strong linear dependency with respect to $k$ and $bc$ as illustrated in Fig.~\ref{fig:Figure_3}(c, d). This is further underlined by correlation coefficients of 0.987 (0.991) of the conditional means with $k\, \left( bc \right)$. Thus, nodes with high $k$ and $bc$, namely the hubs in the random scale-free network, can be classified as its \emph{slow} nodes. Perturbations to a more central node of a scale-free network (operating in the synchronized state) can easily spread to other nodes of the network driving them further away from the synchronized state. As a result, a scale-free network operating in synchrony may take longer to resynchronize when its more central nodes are perturbed as opposed to less central ones. This rationale is supported by the positive correlation between the conditional mean $\langle \langle T_{R}^{1} \rangle \mid bc \rangle$ and $bc$. Further, given the strong linear relationship of the conditional mean \SNRT{} with $bc$, a similar dependence for $k$ is to be expected (and vice versa) since random scale-free networks generally exhibit a strong correlation between $k$ and $bc$ of their nodes~\cite{holme2002attack}. However, the relationship of the conditional mean \SNRT{} with $k$ and $bc$ being specifically linear is surprising and revealing the underlying reason requires further investigation.

We now calculate and present the \GRT{} $\langle T_{R} \rangle$ of all members of the considered ensemble of random scale-free networks (Fig.~\ref{fig:Figure_4}(a), black circles). Interestingly, we observe that unlike for the deterministic scale-free network, the overall time scale of synchronization in the different network realizations of its random counterpart differs markedly from the maximum \SNRT{} (red crosses) of the respective realization. To further study this finding, for each network realization we compute the average of the $\langle T_{R}^{1} \rangle$ values of all its $N = 81$ nodes and denote it by $\langle \langle T_{R}^{1} \rangle \rangle$. Notably, the $\langle T_{R} \rangle$ value of every network realization appears closely related to $\langle \langle T_{R}^{1} \rangle \rangle$ (blue crosses) as illustrated in Fig.~\ref{fig:Figure_4}(a). This is also corroborated by a correlation coefficient of 0.991 between $\langle T_{R} \rangle$ and $\langle \langle T_{R}^{1} \rangle \rangle$.

We further calculate and present in Fig.~\ref{fig:Figure_4}(b) the maximum betweenness centrality $bc_{max}$ of all nodes of each network realization and investigate its relationship with the \GRT{} $\langle T_{R} \rangle$ of the respective realization. As mentioned earlier, perturbing the node with $bc_{max}$ in a scale-free network (operating in the synchronized state) may lead to a particularly large relaxation time to the synchronized state. Thus, the higher the maximum betweenness centrality of a scale-free network, the higher is the \GRT{} of the system, which is underlined by the positive correlation coefficient of 0.882 between $\langle T_{R} \rangle$ and $bc_{max}$ in Fig.~\ref{fig:Figure_4}(b).

The average path length $\mathcal{L}$ of a network is defined as the mean value of the shortest path length between all possible pairs of vertices~\cite{newman2010networks}. Thus, $\mathcal{L} = \frac{1}{N \left( N - 1 \right)} \sum\limits_{i \neq j} \ell \left( i, j \right)$ where $\ell \left( i, j \right)$ is the length of the shortest path between nodes $i$ and $j$ of the $N$-node network~\cite{newman2010networks}. The dependence of the \GRT{} $\langle T_{R} \rangle$ of each network realization on its average path length $\left( \mathcal{L} \right)$ is presented in Fig.~\ref{fig:Figure_4}(c). We observe that $\langle T_{R} \rangle$ exhibits a negative correlation coefficient of -0.658 with respect to $\mathcal{L}$, i.e., random scale-free networks with shorter characteristic path lengths synchronize slower. This result is compatible with the fact that random scale-free networks with longer characteristic path lengths have been previously shown to promote synchronizability and vice versa~\cite{nishikawa2003heterogeneity}. The underlying heuristic picture is that a small $\mathcal{L}$ in such networks corresponds to a large amount of traffic passing through the few `central' nodes connected to each other which facilitate communication between the much larger population of the other oscillators. This may lead to destructive interference of the different signals passing through such nodes. Subsequently, there may not be significant overall communication between the different oscillators of the network, thereby culminating in its reduced synchronizability~\cite{nishikawa2003heterogeneity}.


\subsection{\label{sec:PG_SOKM_UK}Power grid of the United Kingdom}

\begin{figure}
\begin{center}
\includegraphics[height=9.0cm, width=7.5cm]{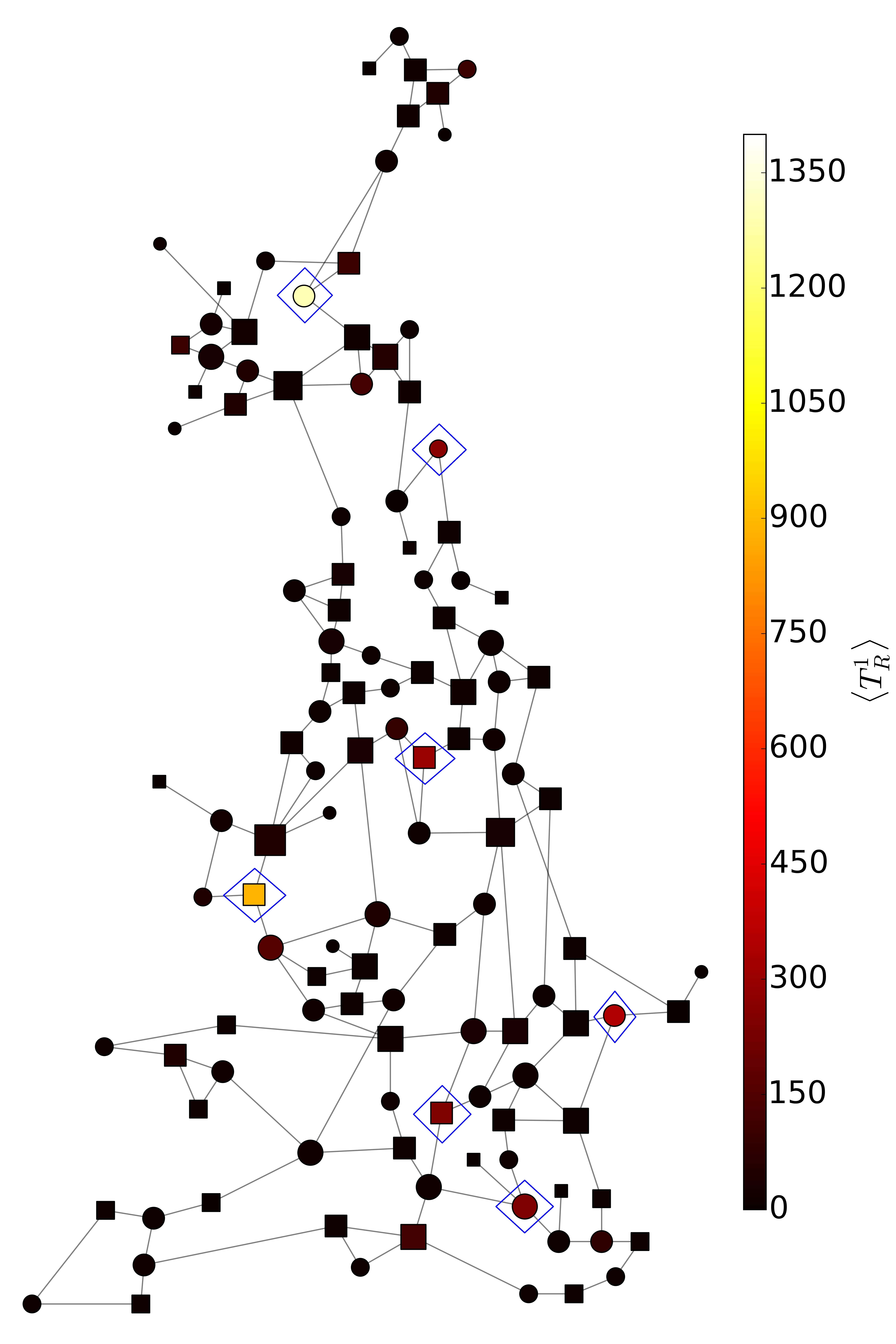}
\caption{\label{fig:Figure_5}(Color online) Network topology of the power transmission grid of the United Kingdom (comprising $N = 120$ nodes) with second-order Kuramoto-type nodal dynamics. Circular nodes denote net generators while square nodes are net consumers. The size of each node is proportional to its degree, and its color corresponds to the $\langle T_{R}^{1} \rangle$ value of the respective node. The 7 nodes further encircled by blue diamonds comprise the \emph{slow} nodes of the grid in our simplified model.}
\end{center}
\end{figure}

\begin{figure*}
\begin{center}
\includegraphics[height=8.0cm, width=18.0cm]{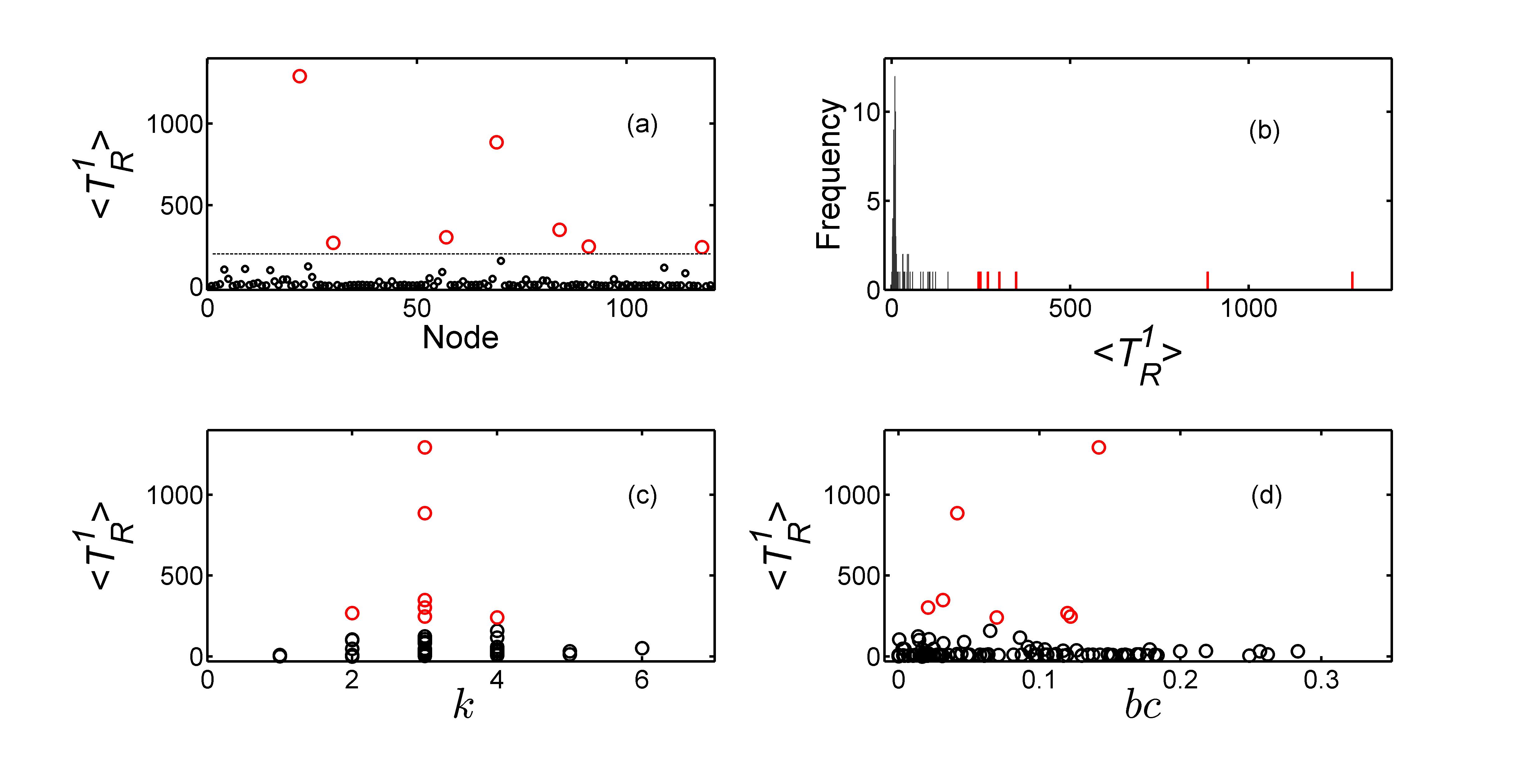}
\caption{\label{fig:Figure_6}(Color online) (a) \SNRT{} $\langle T_{R}^{1} \rangle$ of all the $N = 120$ nodes of the power grid of the United Kingdom with second-order Kuramoto-type nodal dynamics. (b) Histogram of $\langle T_{R}^{1} \rangle$ of all the $N = 120$ nodes. (c, d) Dependence of $\langle T_{R}^{1} \rangle$ on (c) degree ($k$) and (d) betweenness centrality ($bc$) of the nodes. The \emph{fast} nodes of the grid with $\langle T_{R}^{1} \rangle \le 200$ are shown in black while the \emph{slow} nodes having $\langle T_{R}^{1} \rangle > 200$ are marked in red.}
\end{center}
\end{figure*}

As a final more realistic example, we consider a conceptual model of the power transmission grid of the United Kingdom with second-order Kuramoto-type nodal dynamics~\cite{rodrigues2016kuramoto}. The network consists of $N = 120$ nodes and 165 transmission lines (as illustrated in Fig.~\ref{fig:Figure_5}) with topological properties much different from those of a scale-free network. The dynamical equations of the system (in analogy with Eq.~(\ref{eq:DE_Network})) read~\cite{mitra2017multiple}
\begin{equation} \label{eq:PG_SOKM_UK_DE}
\begin{split}
\dot{\theta}_{i} & = \omega_{i},\\
\dot{\omega}_{i} & = - \alpha \omega_{i} + P_{i} + \epsilon \sum\limits_{j = 1}^{N} A_{ij} \sin \left( \theta_{j} - \theta_{i} \right),
\end{split}
\end{equation}
where $\theta_{i}$, $\omega_{i}$, $\alpha$ and $P_{i}$ denote the phase, frequency, electromechanical damping constant and net power input of the $i\textsuperscript{th}$ oscillator, respectively. Furthermore, we randomly choose $\frac{N}{2}$ net generators and $\frac{N}{2}$ net consumers with $P_{i} = +P_{0}$ and $P_{i} = -P_{0}$, respectively~\cite{menck2014dead}. We use the parameter values of $\alpha = 0.1$, $P_{0} = 1.0$ and $\epsilon = 9.0$ for obtaining the results described below.

We again consider the synchronized state, which corresponds to all oscillators having constant phases $\tilde{\theta^{i}}$ and frequencies $\tilde{\omega^{i}} = 0$, as the \DOS{} of the grid. We select $I_{C} = 1000$ trials for calculating the \SNRT{} values of the network. The $\langle T_{R}^{1} \rangle$ values (for $\delta = 10^{-4}$) of all the $N = 120$ nodes are shown in Fig.~\ref{fig:Figure_6}(a) and Fig.~\ref{fig:Figure_6}(b) displays a histogram of all $\langle T_{R}^{1} \rangle$ values. Interestingly, we observe from Fig.~\ref{fig:Figure_6}(a, b) that 113 nodes have low values of \SNRT{} ($\langle T_{R}^{1} \rangle \le 200$), which are shown in black in Fig.~\ref{fig:Figure_6}. However, we also observe 7 \emph{slow} nodes that exhibit substantially higher values ($\langle T_{R}^{1} \rangle > 200$), which are marked in red in Fig.~\ref{fig:Figure_6}. Therefore, (individually or collectively) perturbing any of these 7 nodes of the network will result in dysfunction of the grid and a significantly longer time until the system retaliates to the synchronized state. In turn, it is recommended to control or safeguard these 7 specific nodes of the network to avoid long awaiting time for the system to return to the synchronized state in the face of random perturbations. The choice of the boundary at $\langle T_{R}^{1} \rangle = 200$ for distinguishing between the \emph{fast} and \emph{slow} nodes is motivated by the fact that we observe a first substantial gap in the histogram in Fig.~\ref{fig:Figure_6}(b) around the aforementioned value. We also find similar results from a cluster analysis of the $\langle T_{R}^{1} \rangle$ values of the network. These 7 nodes are not found to exhibit any specific topological features leading to their relatively higher respective $\langle T_{R}^{1} \rangle$ values. Further investigations analyzing these results may provide potentially important insights in this regard.

We emphasize that Erd\H{o}s-R\'{e}nyi random networks~\cite{newman2010networks} of R\"{o}ssler oscillators are found to exhibit similar distributions of $\langle T_{R}^{1} \rangle$ values as above; the corresponding results are described in~\ref{sec:Appendix_ERRN_RO}. Figure~\ref{fig:Figure_6}(c, d) illustrates the values of $\langle T_{R}^{1} \rangle$ in comparison with $k$ and $bc$, respectively. The correlation coefficients of $\langle T_{R}^{1} \rangle$ with $k$ and $bc$ are 0.102 and 0.061, respectively, ruling out the existence of a systematic dependence between $\langle T_{R}^{1} \rangle$ and $k$ or $bc$. Figure~\ref{fig:Figure_5} displays the network topology together with the individual $\langle T_{R}^{1} \rangle$ values in analogy with Fig.~\ref{fig:Figure_2} for the deterministic scale-free network of R\"{o}ssler oscillators.


\section{\label{sec:Conclusions}Conclusions}

Complex systems modelled as networks of interacting dynamical units are ubiquitous and often exhibit multiple stable states. Maintaining operation of such systems in the desired stable state (which often concurs with the synchronized state of the network) is vital to their functionality. Subsequently, this has generated a lot of attention in studying stability of the desired operational state (\DOS{}) in such coupled dynamical systems. However, given that the \DOS{} is stable in principle, it is equally important that the system relaxes back to the same as quickly as possible, following a random perturbation to a particular node of the network. We have addressed this issue here by proposing the general framework of \emph{single-node recovery time} (\SNRT{}) which relates to the time taken by the system operating in the \DOS{} to return to the same, following a non-infinitesimal perturbation to the dynamical state of the respective node. It is important to note that we did not address the problem of driving the perturbed system to the \DOS{}. Instead, we aimed at unveiling the different relative time scales underlying the transient dynamics of individual nodes of the network during its relaxation to the \DOS{}, in order to identify specific nodes which when perturbed lead to significantly enlarged \RT{}. We thus recommend taking precautionary measures of safeguarding primarily these nodes of the network from external perturbations.

Importantly, the proposed machinery can be utilized in revealing relationships between topological features of nodes and their respective \SNRT{} values and in turn, the global relaxation time (\GRT{}) of the overall network. Further, we have suggested the association of \SNRT{} with the concept of engineering resilience in quantifying the resilience of such networked dynamical systems. Finally, we have applied the framework of \SNRT{} to deterministic and random scale-free networks of R\"{o}ssler oscillators and a conceptual model of the power grid of the United Kingdom with second-order Kuramoto-type nodal dynamics.

We have presented here the framework of \SNRT{} (and associated illustrations) in the special context of networks of identical oscillators with continuous time dynamics (Eq.~(\ref{eq:DE_Network})) exhibiting bistability on account of coexisting synchronized and desynchronized regimes. However, the framework is generally applicable to any networked (continuous or discrete time) dynamical system with non-identical nodes and multiple coexisting states. Thus, future work on \SNRT{} could comprise its extension and application to networks of non-identical nodes and/or exhibiting more complex patterns of multistability. Further development on \SNRT{} could comprise its generalization to a framework of \emph{multiple-node recovery time}, similar to recent work in the context of basin stability~\cite{mitra2017multiple}.

Regarding a potential field of application, we emphasize that time-delays arise frequently in the inherent dynamics of individual oscillators and in their interactions on complex networks~\cite{mitra2014dynamical}. Therefore, another interesting endeavour could constitute incorporating time-delays in networked dynamical systems and investigating their influence on \SNRT{} and \GRT{} of the network. Finally, complex systems comprising oscillators coupled on prototypical network types such as Watts-Strogatz, multilayer, interdependent, etc. are open to applications of \SNRT{}. These ventures could further unravel interesting relationships between \SNRT{} and topological features of the aforementioned networks.


\section*{Acknowledgments}

CM and RVD have been supported by the German Federal Ministry of Education and Research (BMBF) via the Young Investigators Group CoSy-CC\textsuperscript{2} (grant no. 01LN1306A). TK, JK \& RVD acknowledge support from the IRTG 1740/TRP 2011/50151-0, funded by the DFG/FAPESP. The authors gratefully acknowledge the European Regional Development Fund (ERDF), the German Federal Ministry of Education and Research (BMBF) and the Land Brandenburg for supporting this project by providing resources on the high performance computer system at the Potsdam Institute for Climate Impact Research.


\section*{References}

\bibliographystyle{unsrt}
\bibliography{References}

\begin{thebibliography}{10}

\bibitem{strogatz2001exploring}
Steven~H Strogatz.
\newblock Exploring complex networks.
\newblock {\em Nature}, 410(6825):268--276, 2001.

\bibitem{albert2002statistical}
R{\'e}ka Albert and Albert-L{\'a}szl{\'o} Barab{\'a}si.
\newblock Statistical mechanics of complex networks.
\newblock {\em Reviews of Modern Physics}, 74(1):47, 2002.

\bibitem{dorogovtsev2002evolution}
Sergey~N Dorogovtsev and Jose~FF Mendes.
\newblock Evolution of networks.
\newblock {\em Advances in Physics}, 51(4):1079--1187, 2002.

\bibitem{newman2003structure}
Mark~EJ Newman.
\newblock The structure and function of complex networks.
\newblock {\em SIAM Review}, 45(2):167--256, 2003.

\bibitem{boccaletti2006complex}
Stefano Boccaletti, Vito Latora, Yamir Moreno, Martin Chavez, and D-U Hwang.
\newblock Complex networks: Structure and dynamics.
\newblock {\em Physics Reports}, 424(4):175--308, 2006.

\bibitem{newman2010networks}
Mark Newman.
\newblock {\em Networks: An Introduction}.
\newblock Oxford University Press, New York, 2010.

\bibitem{pikovsky2003synchronization}
Arkady Pikovsky, Michael Rosenblum, and J{\"u}rgen Kurths.
\newblock {\em Synchronization: A Universal Concept in Nonlinear Sciences},
  volume~12.
\newblock Cambridge University Press, Cambridge, 2003.

\bibitem{menck2014dead}
Peter~J Menck, Jobst Heitzig, J{\"u}rgen Kurths, and Hans~Joachim Schellnhuber.
\newblock How dead ends undermine power grid stability.
\newblock {\em Nature Communications}, 5(3969), 2014.

\bibitem{menck2013basin}
Peter~J Menck, Jobst Heitzig, Norbert Marwan, and J{\"u}rgen Kurths.
\newblock How basin stability complements the linear-stability paradigm.
\newblock {\em Nature Physics}, 9(2):89--92, 2013.

\bibitem{mitra2017multiple}
Chiranjit Mitra, Anshul Choudhary, Sudeshna Sinha, J{\"u}rgen Kurths, and
  Reik~V Donner.
\newblock Multiple-node basin stability in complex dynamical networks.
\newblock {\em Physical Review E}, 95(3):032317, 2017.

\bibitem{arenas2008synchronization}
Alex Arenas, Albert D{\'\i}az-Guilera, J{\"u}rgen Kurths, Yamir Moreno, and
  Changsong Zhou.
\newblock Synchronization in complex networks.
\newblock {\em Physics Reports}, 469(3):93--153, 2008.

\bibitem{pecora1998master}
Louis~M Pecora and Thomas~L Carroll.
\newblock Master stability functions for synchronized coupled systems.
\newblock {\em Physical Review Letters}, 80(10):2109, 1998.

\bibitem{hellmann2016survivability}
Frank Hellmann, Paul Schultz, Carsten Grabow, Jobst Heitzig, and J{\"u}rgen
  Kurths.
\newblock Survivability of deterministic dynamical systems.
\newblock {\em Scientific Reports}, 6(29654), 2016.

\bibitem{zumdieck2004long}
Alexander Zumdieck, Marc Timme, Theo Geisel, and Fred Wolf.
\newblock Long chaotic transients in complex networks.
\newblock {\em Physical Review Letters}, 93(24):244103, 2004.

\bibitem{timme2004topological}
Marc Timme, Fred Wolf, and Theo Geisel.
\newblock Topological speed limits to network synchronization.
\newblock {\em Physical Review Letters}, 92(7):074101, 2004.

\bibitem{timme2006speed}
Marc Timme, Theo Geisel, and Fred Wolf.
\newblock Speed of synchronization in complex networks of neural oscillators:
  analytic results based on random matrix theory.
\newblock {\em Chaos}, 16(1):015108, 2006.

\bibitem{qi2008fast}
GX~Qi, HB~Huang, L~Chen, HJ~Wang, and CK~Shen.
\newblock Fast synchronization in neuronal networks.
\newblock {\em EPL (Europhysics Letters)}, 82(3):38003, 2008.

\bibitem{qi2008predicting}
GX~Qi, HB~Huang, CK~Shen, HJ~Wang, and L~Chen.
\newblock Predicting the synchronization time in coupled-map networks.
\newblock {\em Physical Review E}, 77(5):056205, 2008.

\bibitem{son2008relaxation}
Seung-Woo Son, Hawoong Jeong, and Hyunsuk Hong.
\newblock Relaxation of synchronization on complex networks.
\newblock {\em Physical Review E}, 78(1):016106, 2008.

\bibitem{zillmer2009very}
R{\"u}diger Zillmer, Nicolas Brunel, and David Hansel.
\newblock Very long transients, irregular firing, and chaotic dynamics in
  networks of randomly connected inhibitory integrate-and-fire neurons.
\newblock {\em Physical Review E}, 79(3):031909, 2009.

\bibitem{granada2009achieve}
Adri{\'a}n~E Granada and Hanspeter Herzel.
\newblock How to achieve fast entrainment? the timescale to synchronization.
\newblock {\em PLoS One}, 4(9):e7057, 2009.

\bibitem{grabow2010small}
Carsten Grabow, Steven~M Hill, Stefan Grosskinsky, and Marc Timme.
\newblock Do small worlds synchronize fastest?
\newblock {\em EPL (Europhysics Letters)}, 90(4):48002, 2010.

\bibitem{grabow2011speed}
Carsten Grabow, Stefan Grosskinsky, and Marc Timme.
\newblock Speed of complex network synchronization.
\newblock {\em The European Physical Journal B}, 84(4):613--626, 2011.

\bibitem{wang2016synchronous}
Sheng-Jun Wang, Ru-Hai Du, Tao Jin, Xing-Sen Wu, and Shi-Xian Qu.
\newblock Synchronous slowing down in coupled logistic maps via random network
  topology.
\newblock {\em Scientific Reports}, 6, 2016.

\bibitem{kittel2016timing}
Tim Kittel, Jobst Heitzig, Kevin Webster, and J{\"u}rgen Kurths.
\newblock Timing of transients: Quantifying reaching times and transient
  behavior in complex systems.
\newblock {\em arXiv preprint arXiv:1611.07565}, 2016.

\bibitem{fries2005mechanism}
Pascal Fries.
\newblock A mechanism for cognitive dynamics: neuronal communication through
  neuronal coherence.
\newblock {\em Trends in Cognitive Sciences}, 9(10):474--480, 2005.

\bibitem{fell2011role}
Juergen Fell and Nikolai Axmacher.
\newblock The role of phase synchronization in memory processes.
\newblock {\em Nature Reviews Neuroscience}, 12(2):105--118, 2011.

\bibitem{nishikawa2003heterogeneity}
Takashi Nishikawa, Adilson~E Motter, Ying-Cheng Lai, and Frank~C Hoppensteadt.
\newblock Heterogeneity in oscillator networks: Are smaller worlds easier to
  synchronize?
\newblock {\em Physical Review Letters}, 91(1):014101, 2003.

\bibitem{watts1998collective}
Duncan~J Watts and Steven~H Strogatz.
\newblock Collective dynamics of `small-world' networks.
\newblock {\em Nature}, 393(6684):440--442, 1998.

\bibitem{barahona2002synchronization}
Mauricio Barahona and Louis~M Pecora.
\newblock Synchronization in small-world systems.
\newblock {\em Physical Review Letters}, 89(5):054101, 2002.

\bibitem{holling1973resilience}
Crawford~S Holling.
\newblock Resilience and stability of ecological systems.
\newblock {\em Annual Review of Ecology and Systematics}, 4:1--23, 1973.

\bibitem{holling1996engineering}
Crawford~Stanley Holling.
\newblock {\em Engineering Resilience versus Ecological Resilience}.
\newblock Engineering Within Ecological Constraints. National Academy Press,
  Washington DC, 1996.

\bibitem{pimm1984complexity}
Stuart~L Pimm.
\newblock The complexity and stability of ecosystems.
\newblock {\em Nature}, 307(5949):321--326, 1984.

\bibitem{gunderson2000ecological}
Lance~H Gunderson.
\newblock Ecological resilience--in theory and application.
\newblock {\em Annual Review of Ecology and Systematics}, 31:425--439, 2000.

\bibitem{mitra2015integrative}
Chiranjit Mitra, J{\"u}rgen Kurths, and Reik~V Donner.
\newblock An integrative quantifier of multistability in complex systems based
  on ecological resilience.
\newblock {\em Scientific Reports}, 5(16196), 2015.

\bibitem{walker2004resilience}
Brian Walker, Crawford~S Holling, Stephen~R Carpenter, and Ann Kinzig.
\newblock Resilience, adaptability and transformability in social--ecological
  systems.
\newblock {\em Ecology and Society}, 9(2):5, 2004.

\bibitem{gunderson2012foundations}
Lance~H Gunderson, Craig~Reece Allen, and Crawford~S Holling.
\newblock {\em Foundations of Ecological Resilience}.
\newblock Island Press, Washington, 2012.

\bibitem{rossler1976equation}
Otto~E R{\"o}ssler.
\newblock An equation for continuous chaos.
\newblock {\em Physics Letters A}, 57(5):397--398, 1976.

\bibitem{barabasi2001deterministic}
Albert-L{\'a}szl{\'o} Barab{\'a}si, Erzsebet Ravasz, and Tamas Vicsek.
\newblock Deterministic scale-free networks.
\newblock {\em Physica A: Statistical Mechanics and its Applications},
  299(3):559--564, 2001.

\bibitem{note_choice_of_delta}
{\em We suggest choosing a value of $\delta$ depending upon the system of
  interest. This choice should be made to ensure that the system comes
  sufficiently close to the desired attractor $\mathcal{A}$ as well as being
  computationally efficient. For example, we choose a value of $\delta =
  10^{-4}$ for the deterministic scale-free network, whereas $\delta = 10^{-6}$
  for the random scale-free network. This is because in the former case, the
  system generally takes a longer time to enter the $\delta$-environment around
  $\mathcal{A}$. For a further discussion on the estimation of $T_{RR}$, we
  refer the reader to \cite{kittel2016timing}.}

\bibitem{barabasi1999emergence}
Albert-L{\'a}szl{\'o} Barab{\'a}si and R{\'e}ka Albert.
\newblock Emergence of scaling in random networks.
\newblock {\em Science}, 286(5439):509--512, 1999.

\bibitem{holme2002attack}
Petter Holme, Beom~Jun Kim, Chang~No Yoon, and Seung~Kee Han.
\newblock Attack vulnerability of complex networks.
\newblock {\em Physical Review E}, 65(5):056109, 2002.

\bibitem{rodrigues2016kuramoto}
Francisco~A Rodrigues, Thomas K~DM Peron, Peng Ji, and J{\"u}rgen Kurths.
\newblock The kuramoto model in complex networks.
\newblock {\em Physics Reports}, 610:1--98, 2016.

\bibitem{mitra2014dynamical}
Chiranjit Mitra, G~Ambika, and Soumitro Banerjee.
\newblock Dynamical behaviors in time-delay systems with delayed feedback and
  digitized coupling.
\newblock {\em Chaos, Solitons \& Fractals}, 69:188--200, 2014.

\end{thebibliography}


\appendix


\section{\label{sec:Appendix_x_ref}On the choice of the reference trajectory}

We elaborate here on the existence of a reference state such that the condition in Eq.~(\ref{eq:x_ref}) is fulfilled. For any arbitrary $\mathbf{x}_{ref}$ we have the corresponding $T_{RR}$ function, and hence $\langle T_{R}^{1}(i) \rangle$ as well. Now, we can take a new $\mathbf{x}^{\prime}_{ref} = \varphi(-t, \mathbf{x}_{ref})$ where $\varphi(-t, \cdot)$ is the time-evolution operator shifting a state for the time $t$ backwards along the flow and $t = \langle T_{R}^{1} \rangle_{min}$. Using $\mathbf{x}^{\prime}_{ref}$ we have a corresponding $T^{\prime}_{RR}$ function and $\langle T_{R}^{1 \prime}(i) \rangle$. In particular, $\langle T_{R}^{1 \prime} \rangle_{min} = 0$ holds by construction. So, taking $\mathbf{x}^{\prime}_{ref}$ as the reference state fulfils Eq.~(\ref{eq:x_ref}).


\section{\label{sec:Appendix_ERRN_RO}Erd\H{o}s-R\'{e}nyi random networks of R\"{o}ssler oscillators}

\begin{figure*} [h]
\begin{center}
\includegraphics[height=8.0cm, width=12.0cm]{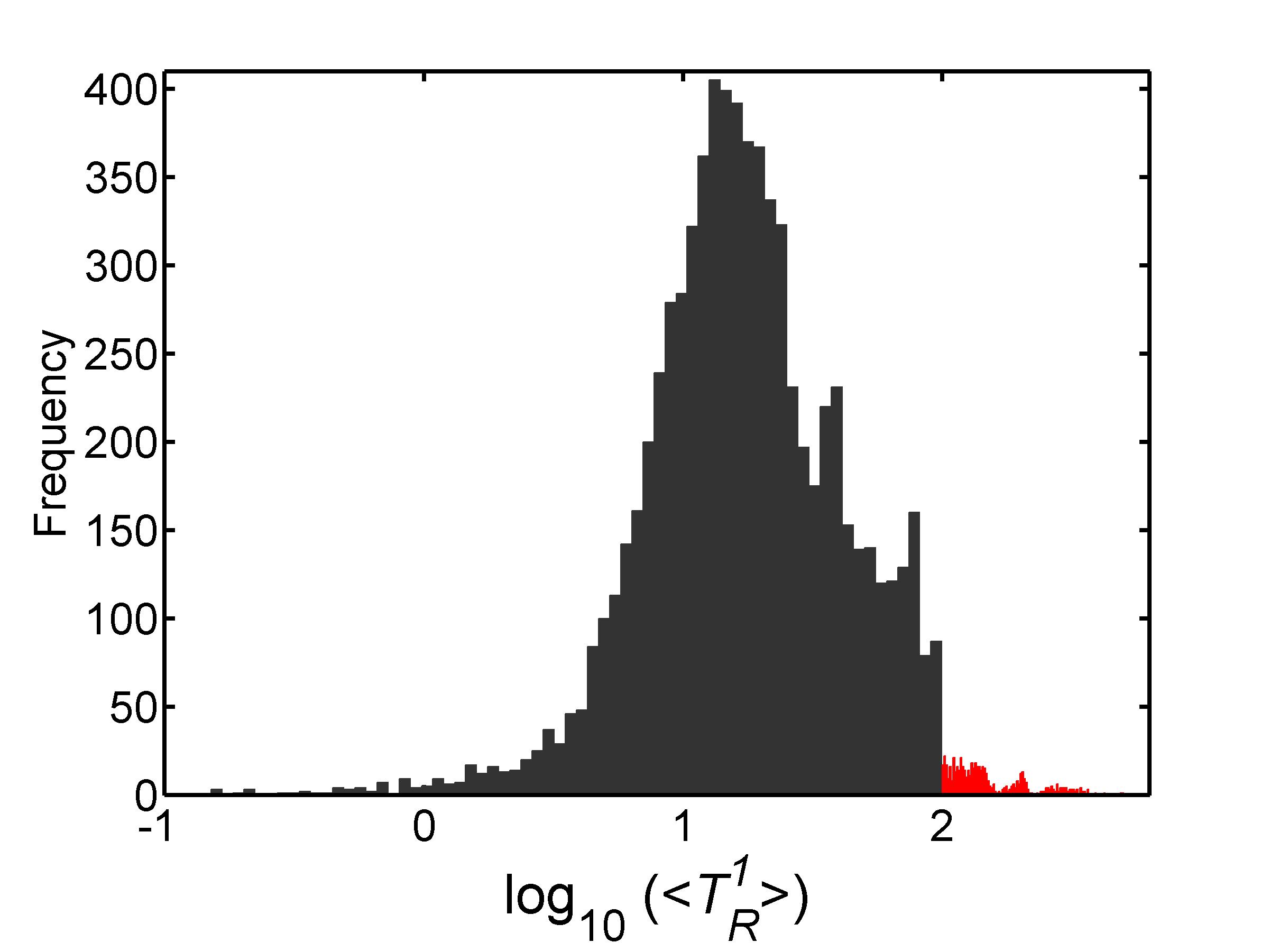}
\caption{\label{fig:Figure_B1}(Color online) \SNRT{} $\langle T_{R}^{1} \rangle$ (on $\log_{10}$ scale) of all nodes of the considered ensemble of Erd\H{o}s-R\'{e}nyi random networks. The \emph{fast} nodes of the ensemble with $\langle T_{R}^{1} \rangle \le 100$ are shown in black while the \emph{slow} nodes having $\langle T_{R}^{1} \rangle > 100$ are marked in red.}
\end{center}
\end{figure*}

Here, we consider an ensemble of 100 Erd\H{o}s-R\'{e}nyi random networks~\cite{newman2010networks} of $N = 81$ R\"{o}ssler oscillators each, again with the same parameter values as for the deterministic scale-free network (Sec.~\ref{sec:DSFN_RO}). We consider a probability $p = 0.04$ of a connection between any pair of vertices of a network, resulting in a total of 130 edges in each realization. For $\delta = 10^{-6}$, we calculate and present the distribution of $\langle T_{R}^{1} \rangle$ (on $\log_{10}$ scale) values of all nodes of the considered ensemble of Erd\H{o}s-R\'{e}nyi random networks in Fig.~\ref{fig:Figure_B1}. It is evident from the distribution that most nodes have rather low values of $\langle T_{R}^{1} \rangle$ ($\le 100$), which comprise the \emph{fast} nodes of the respective network. However, we also observe the existence of very few \emph{slow} nodes which exhibit much higher $\langle T_{R}^{1} \rangle$ ($> 100$) values. The $\langle T_{R}^{1} \rangle$ values again do not exhibit any strong linear relationship with $k$ ($bc$), as demonstrated by the correlation coefficient of 0.743 (0.36).

\end{document}